\documentclass[titlepage,twoside,12pt]{article}
\usepackage{amssymb}
\usepackage{amsfonts}
\begin{document}

{\bf \Large Covariance and Frames of Reference} \\ \\

{\bf Elem\'{e}r E Rosinger} \\
Department of Mathematics \\
and Applied Mathematics \\
University of Pretoria \\
Pretoria \\
0002 South Africa \\
eerosinger@hotmail.com \\ \\

{\bf Abstract} \\

A transition of focus from {\it state space} to {\it frames of reference} and their {\it
transformations} is argued as being the appropriate setup for ensuring the {\it covariance} of
physical laws. Such an approach can not only simplify and clarify aspects of General
Relativity, but can possibly help in the development of a Grand Unified Theory as well. \\ \\

\hspace{3cm} "Physics may be not only too important, but also

\hspace{3cm} too deep, in order to be left alone to usual physical

\hspace{3cm} type thinking and arguments." \\ \\

{\bf 1. Frames of Reference versus State Space} \\

{\bf 1.1. State Spaces} \\

{\bf Classical.} At least since H Poincar\'{e}, we are accustomed to associate with the
dynamics of a given physical system a corresponding {\it state space} which is the set of all
possible points, each of which uniquely describing one of the possible situations the
respective system can be. Consequently, a state space can be finite or infinite dimensional. \\

For instance, the dynamics of $n \geq 1$ material points in usual 3-dimensional Euclidian
space $\mathbb{R}^{\,3}$ can be described by $6n$-dimensional vectors which give their
individual positions and momenta. Thus the corresponding state space is $\mathbb{R}^{\,6n}$.
In case of a sufficiently laminar or smooth flow of a fluid in a domain $\Omega \subseteq
\mathbb{R}^{\,3}$, the state space can be taken as ${\cal C}^2(\Omega,\mathbb{R})$, each
function $u : \Omega \longrightarrow \mathbb{R},~ u \in {\cal C}^2(\Omega,\mathbb{R})$
describing the field of velocities at a certain moment in time. \\
In such classical situations time need {\it not} be included from the very beginning into the
state space. Instead, it can be introduced as an outside and additional component. Namely,
time is modelled by the real line $\mathbb{R}$, and then enters into consideration by a {\it
cartesian product} with the state space. For instance, in the first case above, we shall have
$\mathbb{R} \times \mathbb{R}^{\,6n}$, while in the second case $\mathbb{R} \times {\cal
C}^2(\Omega,\mathbb{R})$. \\

{\bf Quantum.} Such an approach remains valid also in the non-relativistic quantum mechanics
of a finite number of particles. Indeed, if we have $d \geq 1$ such particles, then the state
space is the Hilbert space ${\cal L}^2(\mathbb{}R^d)$ which contains the wave functions $\psi$
as they are at any given moment of time, while the dynamics happens in $\mathbb{R} \times
{\cal L}^2(\mathbb{}R^d)$. \\

{\bf Dynamics.} The general pattern in the above cases is as follows. There is a {\it state
space} given by a certain set ${\cal S}$, and every specific dynamics of the system is
described by a corresponding curve $C \subseteq {\cal S}$ which is parametrized by {\it time},
according for instance, to a mapping $I \ni t \longmapsto x_t \in C$, where $I \subseteq
\mathbb{R}$ is some time interval. \\
Alternatively, we can consider in such a situation the composite space \\

(1.1)~~~ $ \Sigma = I \times {\cal S} $ \\

in which case every specific dynamics of the system is described by a curve $\Gamma \subseteq
\Sigma$, where $\Gamma = \{~ ( t, x_t ) ~~|~~ t \in I ~\}$. \\

{\bf Relativity.} In Special and General Relativity the situation is different, since from the
start time has to be considered {\it together} with space. In this way, from the beginning the
situation recalls or extends that in (1.1). For instance, for one point particle in Special
Relativity, the state space still has the form (1.1), namely $\Sigma = \mathbb{R} \times
\mathbb{R}^{\,3}$, while considered in General Relativity, the state space $\Sigma$ is a
4-dimensional so called {\it space-time} manifold, resulting from the Einstein equations. \\

{\bf 1.2. Frames of Reference} \\

In Newtonian Mechanics, Maxwell Electro-Magnetism, or for that matter, non-relativistic
Quantum Mechanics of finite systems, the issue of frames of reference and transformations of
such systems comes usually up in limited ways. What happens is that one assumes given a
certain convenient coordinate system which automatically defines a frame of reference, and one
then subsequently tends to remain within it. When a different coordinate system, thus frame of
reference, has to be considered, the respective transformations are performed, including to
the corresponding form of the physical laws of interest. And often such laws may take a
different form, since the transformations used need not always be restricted to those which
leave these laws covariant. \\
In this way, the stress is rather on the state spaces, than on the covariance of the laws
under the transformations encountered. \\

This situation changes significantly starting with Special Relativity where one of the two
fundamental principles is that the laws of Physics are the same in every inertial frame of
reference, and this principle is frequently and essentially used in the development of the
respective theory. \\
It follows that in Special and General Relativity frames of reference and their
transformations are from the very beginning among the building blocks of the respective
theories, thus in fact they acquire a new prominent position versus state spaces. \\

In this regard, it will be argued in the sequel that in pursuing both General Relativity and
Grand Unified Theories, it may be appropriate to focus {\it more} on frames of reference and
their transformations, as well as the requirement that the laws of Physics remain covariant
under the respective transformations, than on the state spaces. \\

{\bf 2. Starting with Frames of Reference and Covariance} \\

With hindsight, we can now realize that in General Relativity by far the most important fact
is that the equations of Physics must be covariant under arbitrary ${\cal
C}^2$-diffeomorphisms. \\
In other words, just as much as from the point of view of Physics as such it is absolutely
irrelevant whether a book on the subject is written in English or in any other sophisticated
enough language, so it must be absolutely irrelevant for their form in which coordinate system
the laws of Physics are written, see arXiv:physics/0505045. \\

As it happens, however, the way A Einstein pursued from Special Relativity to General
Relativity during the decade 1905-1915, or for that matter, the last moment involvement of D
Hilbert, did not have as starting point this argument of very general type of covariance, see
Kaku. \\
Instead, a variety of arguments of physical nature had been involved, some of them leading
nowhere, even if the initial insight of A Einstein related to what is nowadays called the
Equivalence Principle proved to be correct and fundamental. \\

Here it may be important to note that such an approach which brings in an abundant variety of
physical type arguments, often regardless of their effective relevance, is rather typical in
the development of Physics. Yet, such an approach can often lead to unnecessary prolixity, and
thus possible lack of clarity or confusion, and in addition, also to the risk of missing some
really important points. A nontrivial example in this regard is the way quantum physicists,
for instance, tend so often to argue and obtain the celebrated Bell inequalities. Indeed, most
of the rather endless arguments ranging around the Bell inequalities, arguments in which any
amount of physical insights may be marshalled, happen to miss the point that the Bell
inequalities - as inequalities - belong to a family of classical inequalities which were
already known to George Boole back in 1856, see arXiv:quant-ph/0406004. \\
Needless to say, such a situation should serve as a good warning, and not only in Quantum
Mechanics. \\

And then, there are only two issues left to consider :

\begin{itemize}

\item which laws of Physics we talk about,

\item what do we mean by frames of reference.

\end{itemize}

When it comes to frames of reference, however, one usually proceeds quite automatically, and
without much thinking. Namely, one starts with a state space, with corresponding coordinate
systems, and then one associates frames of reference and talks about smooth enough and
invertible transformations of the state space one may happen to use. \\

Therefore, one may have to revise such an approach as follows :

\begin{itemize}

\item first, reconsider deeply enough what we really mean by frames of reference, and what
such systems may indeed be like,

\item then, simply ask what are the covariant laws of Physics in such frames of reference.

\end{itemize}

The advantage of such an approach may be twofold :

\begin{itemize}

\item to simplify the story of General Relativity,

\item to help us in setting up a Grand Unified Theory.

\end{itemize}

As for the second point above, it is indeed obvious that the present rather automatic and
unquestioned approach to the concept of frames of reference is one of the major factors which
lead us to the impossibility of bridging the divide between the "continuum" aspect of General
Relativity, and on the other hand, of the "discrete" aspect of Quantum Theory. \\

In his recent book "Einstein's Cosmos", M Kaku stresses that two very simple but fundamental
physical ideas had led Einstein to his setting up relativity. For Special Relativity he used
his teenage idea of racing along a beam of light, while for General Relativity he used what
came to be called the "Equivalence Principle". Further, M Kaku suggests that the last three
decades of Einstein's life, when he attempted a Grand Unified Theory, did not lead to a
success since he could not find a third such a simple but equally fundamental physical idea to
start with and then build upon. \\
It is also suggested that Special Relativity was considered by Einstein insufficient, since it
did not include gravitation. \\
As for finding a Grand Unified Theory, Einstein kept talking about "marble and wood", of
turning "wood into marble", of creating a theory of "pure marble" … \\

The fact however is that the need not to stop at Special Relativity, and instead go further
for General Relativity has, in terms of frames of reference and covariance, a very simple and
most obvious direct reason, namely :

\begin{quote}

Why should covariance be limited only to Lorentz transformations of frames of reference ?

\end{quote}

In this way, such a simple question, arXiv:physics/0505045, is all alone and all in itself
sufficient in order to get us going from Special Relativity towards General Relativity. And
once one realizes that, one can of course use arguments of Physics in order to set up the
respective theory. \\
However, on the level of motivation for not remaining with Special Relativity, one does not
need to be much concerned about any Physics at all, except to remember that there are far more
general frames of reference and coordinate transformations than the Lorentzian ones. Thus
covariance cannot and should not be limited to what it means in Special Relativity. \\

As for creating a "pure theory of marble", that is, creating a Grand Unified Theory, it may
well happen that an argument which for the time being is equally overlooked is in fact needed.
An argument which, as it happens at present, is replaced by endless and not particularly
effective physical considerations. Namely, and as mentioned above, it may well happen that we
shall first have to work out much more general concept of frames of reference than used in
General Relativity, a concept which can indeed bring properly together the "marble" and the
"wood" ... \\

{\bf 3. Precedents} \\

The idea of abandoning usual state spaces in Physics already has precedents, one of them as
famous as String Theory. \\
In another direction, however along the same intent, categories and toposes have been used,
see for instance Coecke, Isham, Butterfield \& Isham and the references cited there, for some
of the more recent such approaches. \\

\end{document}